\documentclass{article}

\PassOptionsToPackage{numbers, compress}{natbib}

\usepackage[preprint]{neurips_2024}

\usepackage[utf8]{inputenc} 
\usepackage[T1]{fontenc}    
\usepackage{hyperref}       
\usepackage{url}            
\usepackage{booktabs}       
\usepackage{amsfonts}       
\usepackage{nicefrac}       
\usepackage{microtype}      
\usepackage{xcolor}         
\usepackage{amsmath}
\usepackage{graphicx} 
\usepackage{wrapfig}
\usepackage{xcolor}
\usepackage{algorithm}
\usepackage{algpseudocode}
\usepackage{graphicx}
\usepackage{comment}
\definecolor{prompt_gray}{RGB}{220,220,220}
\title{Mixture-of-PageRanks:\\Replacing Long-Context with\\Real-Time, Sparse GraphRAG}

\author{%
  Nick Alonso\\
  Zyphra\\
  \texttt{nick@zyphra.com} \\
  \And
  Beren Millidge \\
  Zyphra \\
  \texttt{beren@zyphra.com} \\
}

\begin{document}

\maketitle

\begin{abstract}
Recent advances have extended the context window of frontier LLMs dramatically, from a few thousand tokens up to millions, enabling entire books and codebases to fit into context. However, the compute costs of inferencing long-context LLMs are massive and often prohibitive in practice. RAG offers an efficient and effective alternative: retrieve and process only the subset of the context most important for the current task. Although promising, recent work applying RAG to long-context tasks has two core limitations: 1) there has been little focus on making the RAG pipeline compute efficient, and 2) such works only test on simple QA tasks, and their performance on more challenging tasks is unclear. To address this, we develop an algorithm based on PageRank, a graph-based retrieval algorithm, which we call mixture-of-PageRanks (MixPR). MixPR uses a mixture of PageRank-based graph-retrieval algorithms implemented using sparse matrices for efficent, cheap retrieval that can deal with a variety of complex tasks. Our MixPR retriever achieves state-of-the-art results across a wide range of long-context benchmark tasks, outperforming both existing RAG methods, specialized retrieval architectures, and long-context LLMs despite being far more compute efficient. Due to using sparse embeddings, our retriever is extremely compute efficient, capable of embedding and retrieving millions of tokens within a few seconds and runs entirely on CPU.\footnote{Thanks for Jonathan Pilault, Vasu Shyam, and the rest of the Zyphra team for comments and discussions on earlier drafts of this paper.}

\end{abstract}
\begin{figure}[h]
\centering
\includegraphics[width=0.95\textwidth]{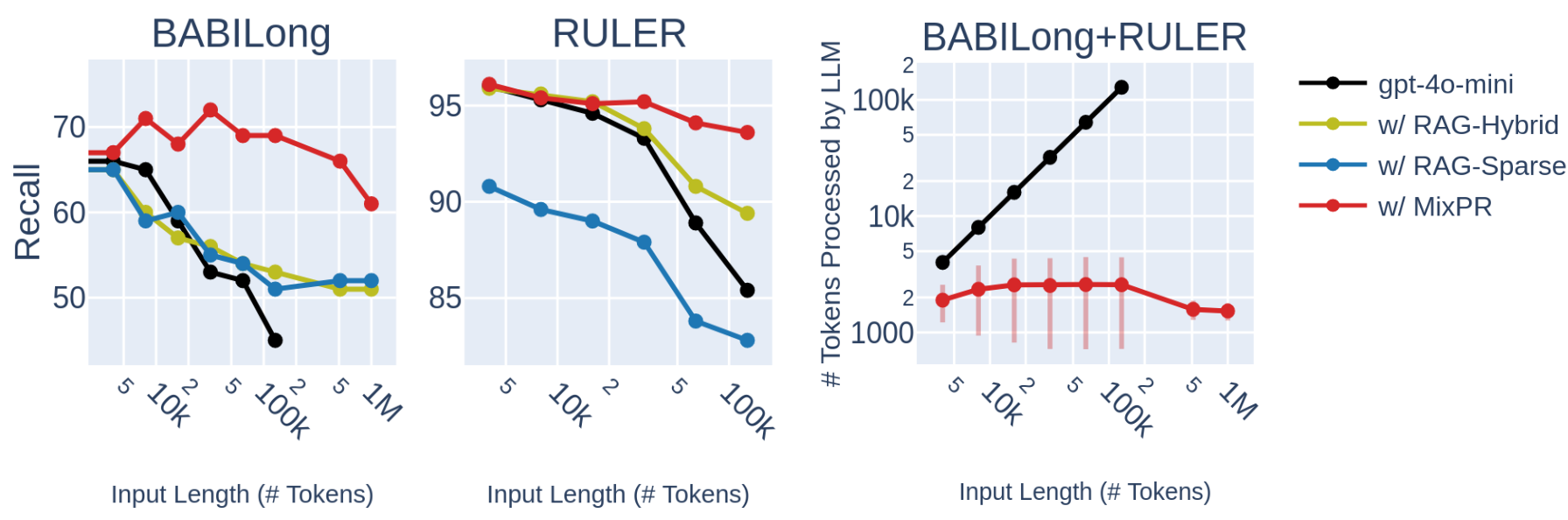}
\caption{Comparison of gpt-4o-mini and RAG systems (@k=100) that use gpt-4o-mini as the generator. (Left) Our MixPR outperforms baseline RAG models and the base LLM on multiple standard long-context benchmarks.(Right) RAG models, like our MixPR, use only a small portion of input text, drastically reducing compute cost of LLM inference.}\label{fig:front_page}
\end{figure}
\section{Introduction}

Frontier long-context LLMs are now capable of processing hundreds of thousands, even millions, of tokens directly in the prompt. This allows LLMs to reason over long documents like entire books, code-bases, and collections of articles.  However, processing long-contexts is extremely computationally expensive due both to the quadratic cost of attention and the memory cost of storing a large KV-cache. This makes inferencing long-context models challenging in practice and often impossible on compute-limited devices. Additionally, compute costs of long-context models also make API calls highly expensive. Moreover, it is unclear whether these increased compute costs are always worth it, given that recent benchmarks testing long-context recall and reasoning show that SOTA LLMs often struggle to perform a variety of tasks for long input lengths (e.g., \cite{hsieh2024ruler, kuratov2024babilong}). 

One significantly cheaper and more extendable alternative to full-context processing is retrieval augmented generation (RAG). The idea here is to first store a long context in a vector database at test-time, then a retrieval algorithm scans the database and returns only the elements that are most relevant to the LLM. By only inputting a small portion of the context, one can drastically reduce the compute cost and latency of LLM inference. Further, it has been commonly observed that LLM performance tends to improve when irrelevant and/or distracting information is removed from (e.g., \cite{levy2024sametask}), suggesting RAG could actually improve performance over full-context processing.

Though promising, recent work developing RAG algorithms for long-context tasks have several limitations: Firstly, these works (e.g., \cite{yu2024defense, zhao2024longrag}) tend to focus on simple QA style tasks, even though many long context tasks (e.g., reasoning, summarization) that frontier LLMs are capable of are much more complex than this. Secondly, previous works do not analyze the computional costs of the retriever which are themselves often significant. RAG in long-context scenarios operates in a different paradigm than standard RAG. For long-context, RAG must generate text chunk embeddings at test-time for each new input, compared to standard RAG which typically assumes the LLM has access to a pre-embedded database, created prior to test-time. While RAG will generally reduce the compute cost of LLM inference compared to long-context LLM inference, the RAG pipeline itself (chunking, embedding, database entry, and retrieval) can often be slow or expensive enough to preclude real-time embedding and retrieval at test time.

In this paper, we go beyond previous work by making significant progress on both of these issues simultaneously. We introduce both a novel retrieval algorithm which can achieve SOTA performance on challenging multi-step retrieval tasks and an efficient implementation that uses sparse embeddings, which can embed and retrieve from millions of tokens in real-time on CPU. Our contributions can be summarized as follows:
\begin{enumerate}
\item We develop a novel retrieval algorithm based on personalized PageRank (PPR) \cite{brinPage1998anatomy}, a graph-based retrieval method. Our algorithm, which we call mixture-of-PageRanks (MixPR), dynamically adjusts a hyper-parameter in PPR based on the query/task type. We provide mathematical intuition and empirical testing supporting our dynamic version of PPR. MixPR is designed to handle a variety of complex tasks requiring different kinds of information from the context.
\item We design a simple method for generating the adjacency matrix in PPR that is much cheaper compared to current SOTA methods for graph-based RAG, while still yielding high-performance. We also develop an implementation of the PPR retrieval algorithm using sparse matrices that is memory efficient and allows for fast retrieval entirely on CPU. 
\item We extensively test our MixPR retriever with multiple LLMs across a variety of long-context tasks, including four long-context benchmarks, each consisting of multiple sub-tasks, totaling 22 tasks including QA, reasoning, and summarization. MixPR-RAG models achieve SOTA or near SOTA among similar models on three of these benchmarks, despite being far more compute efficient than existing SOTA models. 
\item We empirically show our MixPR retriever can process millions of tokens in several seconds or less on a desktop, evidencing its applicability for long-context processing on-device.
\end{enumerate}

\section{Background}

\textbf{Long-Context Tasks.} Long-context benchmarks (e.g., \cite{kamradtNIAH, team2024gemini, hsieh2024ruler, li2024needlebench, zhang2024infbench, levy2024sametask, goldman2024really, karpinska2024one, MagicHashHop, kuratov2024babilong, vodrahalli2024michelangelo, karpinska2024one}) test the ability of language models to retrieve from and reason over vast text inputs. Such text inputs will typically contain a long document (e.g., a book, multiple articles, a code directory, etc.), a query about the document, and, possibly, a zero-shot or few-shot prompt. The model must respond accurately to the query given the document. Such benchmarks are meant to simulate the scenario where a long input is given to the LLM for processing at \textit{test time}. 

\textbf{Two Challenges for the RAG Approach to Long-Contexts Tasks.} RAG models are commonly used for knowledge retrieval settings. However, there are (at least) two reasons why utilizing RAG effectively in long-context tasks is more difficult than knowledge retrieval tasks.

\begin{enumerate}
    \item \textbf{The Challenging Retrieval Problem.} Standard RAG methods, that utilize a nearest neighbor search over a vector database, work well on standard QA and retrieval tasks: the model is queried about some content X, text chunks most related to X are efficiently retrieved, then utilized by the LLM to accurately respond to the query. However, many long-context tasks are more complex than single step retrieval and require linking together diverse elements across the whole input document, e.g., reasoning, reading comprehension, etc. Further, these elements may not be directly related to the content of the query, as we see with summarization tasks, and standard retrieval methods do not account for this.
    \item \textbf{The Compute Cost Problem.} Common RAG algorithms chunk text and embed the chunks in a vector database, usually using a transformer-based embedding model. In the typical RAG setting, i.e., knowledge retrieval, the embedding of a database happens before deployment, then during deployment, the database is repeatedly and efficiently searched as queries are presented. Long context tasks, on the other hand, require processing a query and a long input document (e.g., an entire book along with questions) \textit{at test time}, which means the embedding step, in addition to the retrieval step, must be completed at test time. Since in many applications the user expects an immediate response, the embedding process must be fast to be practically useful. However, the process of embedding a long list of text chunks using standard transformer-based embedders can be slow, especially in compute constrained settings.
\end{enumerate}

Our aim is to go beyond previous work by designing a RAG system that can both perform well on tasks more difficult and complex than single-step retrieval and will do so in a fast, compute efficient way.

\section{Related Works}
We provide a brief review of related works here. In the appendix, we provide a more detailed analysis. 

\textbf{Long-Context LLMs:} LLM context windows have expanded dramatically in recent works, with frontier LLMs achieving contexts of 128k or over 1 million \cite{team2024gemini, anthropic2023model, achiam2023gpt}. These advances have primarily been driven by position interpolation methods (\cite{peng2023yarn, su2024roformer}) which enable models trained at short contexts to be `extended' to handle long contexts. However, long-context models typically are transformer-based, utilizing attention layers which require compute quadratic in the sequence length for inference and large KV caches for generation. These high compute and memory costs render extremely long contexts either very expensive or practically infeasible.

\textbf{Long-Context Benchmarks:} Benchmarks testing the ability of these LLMs to effectively utilize their context are currently being rapidly developed. Early long context benchmarks focused on retrieval tasks \cite{kamradtNIAH, team2024gemini, hsieh2024ruler, li2024needlebench, zhang2024infbench, kuratov2024babilong, hsieh2024ruler, zhang2024infbench}. More recently, there has been an emphasis on testing more complex tasks than mere single step retrieval \cite{levy2024sametask, goldman2024really, karpinska2024one, vodrahalli2024michelangelo}, such as reasoning \cite{hsieh2024ruler, MagicHashHop, kuratov2024babilong, vodrahalli2024michelangelo}, reading comprehension \cite{karpinska2024one}, and summarization \cite{zhang2024infbench}.

\begin{figure}[t]
\centering
\includegraphics[width=0.94\textwidth]{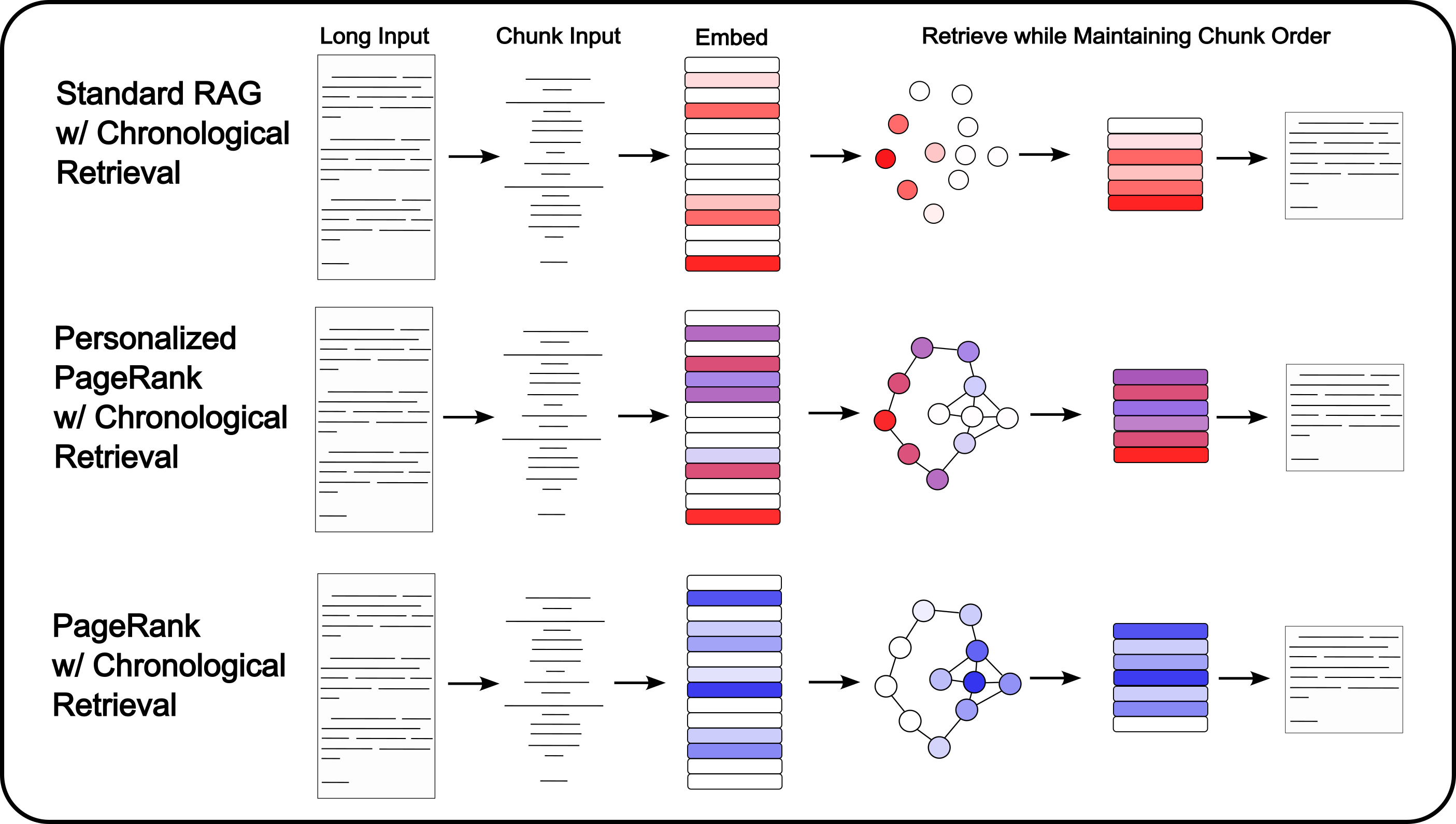}
\caption{\textbf{RAG Methods for Long-Context Tasks.} Previous works \cite{AlonsoMillBabilongBlog, yu2024defense} have shown that chronologically ordering text chunks, rather than rank ordering, is necessary for many long-context tasks. Standard nearest-neighbor RAG (top) retrieve items most similar to the query (query-relatedness is depicted with red embeddings). Our PageRank-based retrievers represent relations between text chunks using a similarity matrix that can be computed cheaply. PageRank (bottom) retrieves items that have the highest importance according to the graph structure (structural importance depicted with blue). Personalized PageRank (middle) balances query-relatedness with structural importance.}
\end{figure}

\textbf{Graph-based RAG for Database Retrieval:} Graph-based RAG systems retrieve from graph structured databases rather than unstructured databases. Edge et al. \cite{edge2024local} show that graph structured databases may be used to better respond to queries about the global properties of the database, rather than local, query-specific properties (see also \cite{guo2024lightrag}). Gutierrez et al. \cite{gutierrez2024hipporag} get SOTA results on multi-hop QA tasks using a graph-structured database and the personalized page rank retrieval algorithm. Other recent promising results suggest graph-based RAG is promising relative to nearest-neighbor search unstructured database retrieval (for review, see\cite{peng2024graph}). These methods typically use a slow procedure involving LLM generation to create the graph database, whereas our algorithm is very compute efficient and uses no neural networks for graph-construction or embedding. Further, unlike these works, our algorithm uses a dynamic method for adjusting a hyper-parameter in the PageRank algorithm based on query type (see below) which enables us to handle both local retrieval tasks such as QA and global tasks such as summarization with a single algorithm.

\textbf{RAG for Long Contexts:} Alonso and Millidge \cite{AlonsoMillBabilongBlog} tested several graph-based RAG systems on the BABILong \cite{kuratov2024babilong}, long-context QA task. They showed chronological ordering of retrieved chunks and graph-based retrieval is necessary to achieve good performance on difficult multi-hop questions. Yu et al. \cite{yu2024defense} subsequently showed that standard RAG systems perform well on two simple long-context QA tasks and provided further evidence chronological ordering of text chunks is important. Zhao et al. \cite{zhao2024longrag} developed a complex RAG system that performs multiple retrieval steps guided by a chain of thought process, and showed it was effective on QA tasks. Though promising, none of these previous works attempt to address the main challenges for RAG-based long-context models, as they do not test on complex reasoning or summarization tasks or attempt to minimize compute costs of retrieval.

\section{The Sparse Mixture of PageRanks Retriever}
We begin with the intuition, similar to Vodrahelli et al \cite{vodrahalli2024michelangelo}, that RAG systems must assign importance to chunks in ways that balance both relatedness to the query as well as the chunk's (query-independent) \textit{structural importance} within the document, e.g., a text chunk describing a main event in a story would have high structural importance even when not directly related to the content of a query (e.g., 'summarize this book'). Combining these two importance measures is akin to computing a posterior distribution via Bayesian inference, which balances the input data-dependent likelihood with the data-independent prior probability. Bayesian inference, importantly, adjusts the balance of the posterior between the prior and likelihood depending on the precision and relevance of the data. Similarly, an effective RAG systems should also be able to adjust the influence of query-relatedness and structural importance depending on the query and task, e.g., \textit{local retrieval tasks} (such as question answering (QA)) require retrieving items that have higher query-relatedness, while \textit{global retrieval tasks} (such as summarization) require retrieving items that have higher structural importance. 

\textbf{PageRank.} We propose modeling the latent structure of a long context document as a graph, which relates text chunks based on their semantic and syntactic similarities. We use the personalized PageRank algorithm \cite{brinPage1998anatomy} (PPR), originally utilized by Google search, to provide a principled method for combining query-relatedness with structural importance, and for controlling the influence of the two during retrieval. The (non-personalized) PageRank (PR) algorithm computes a categorical distribution, $\pi$, over items in a database which is the steady-state of a Markov Chain defined by the Markov adjacency matrix, $A$:
\begin{equation}
    \pi = A\pi,
\end{equation}
where $\pi$ and the columns of $A$ are normalized to sum to one. The PR distribution is proportional to the number of times each node is visited during an infinite long random walk through the graph \cite{lofgren2015efficient}. Nodes with more incoming connections from neighbors and/or more important neighbors tend to have higher PR score. The PR distribution, in this way, captures a pure notion of structural importance of nodes in the graph. 

\textbf{Personalized PageRank.} Personalized PageRank (PPR) is a variant of PageRank that biases the distribution towards a personalization distribution, $p$:
\begin{equation}
\pi = (1 - \alpha) A\pi + \alpha p,
\end{equation}
where $0 < \alpha < 1$ is a scalar weighting term. The PPR distribution models the number of visits to each node during an infinite series of finite random walks, where $p$ is the distribution over start nodes, and whose length is distributed in a way dependent on $\alpha$. In internet search, $p$ is often a data-dependent vector describing the number of times a particular user has visited various websites \cite{brinPage1998anatomy}, thereby personalizing search to that user. PPR thus provides a method of computing a distribution that combines both data-dependent and structural importance. Further, $\alpha$ provides a clear method for controlling how data-related rankings are: as $\alpha \rightarrow 0$ PPR approaches PR and the ranking becomes increasingly data-independent and structurally-dependent. As $\alpha \rightarrow 1$ the distribution becomes increasingly data-dependent and structurally-independent.
\begin{figure}[t]
\centering
\includegraphics[width=0.99\textwidth]{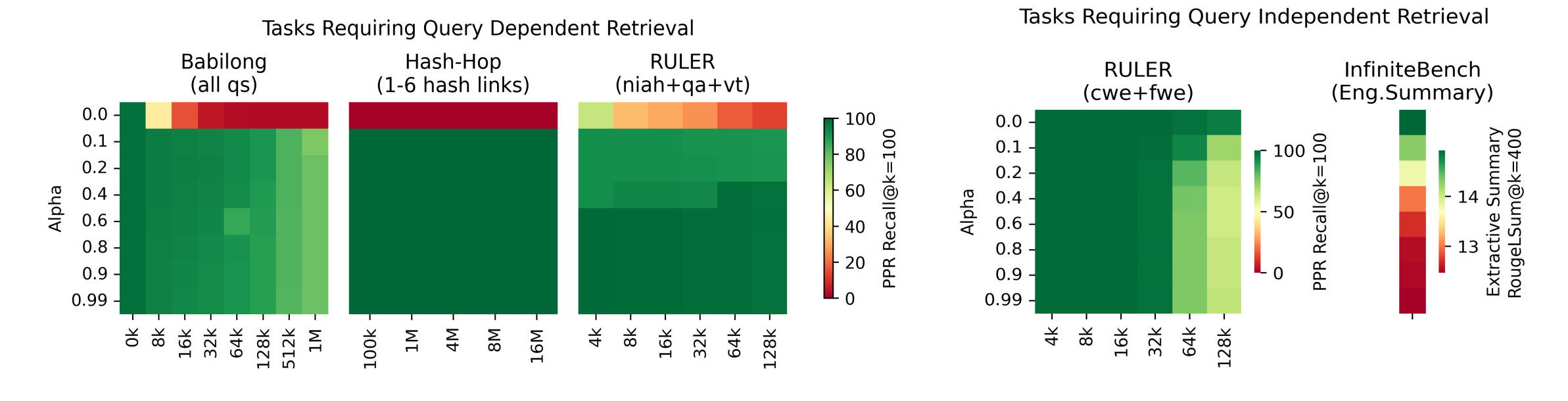}
\caption{\textbf{PPR Retriever Alpha Test.} (Left) The recall accuracy of a PPR retriever at k=100, under various alpha settings, for a set of query-dependent, local retrieval tasks. (Right) The recall performance of the same PPR retrieval at k=100, for query-independent, global retrieval tasks.}\label{fig:alpha_test}
\end{figure}

\textbf{Chunking method.} Any application of PPR to RAG needs a method to chunk text and to compute the $A$, $p$, and $\alpha$ terms of the PPR algorithm. Moreover, the chunking method often heavily affects downstream retrieval performance. In this paper, we use a simple chunking method that first chunks the text into sentences. Any chunks that are greater than $N$ words are then chunked along newline characters. If any chunks are still greater than $N$ words, these are split into smaller chunks of roughly equal length along word boundaries. We set $N=32$ for all experiments.

\textbf{Graph Generation.} Recent SOTA RAG models that use graph search, including PPR, create $A$ using an LLM to output a long series of triplets describing related entities in the text (e.g., \cite{gutierrez2024hipporag, edge2024local}). Since we aim at real-time long-context embedding,  this is much too slow for our purposes. Instead, our algorithm first embeds each chunk into a vector representation (which can, in principle work, with many embedding methods). Each embedding is then L-2 normalized and stored in chronological order in the columns of a matrix $E$. We set $A$ equal to the proximity matrix computed from $E$:
\begin{equation}
A = normalize(E^{\top}E),
\end{equation}
where $normalize(M)$ normalizes columns of $M$ to sum to one.

\textbf{Personalization Vector.} Items in $E$ are in chronological order. This means the last probability in $\pi$ represents the probability of the last/most recent text chunk. We treat the personalization vector $p$ as a one-hot vector whose last value is set to one and all other values are set to zero (or the last two text chunks are set to .5 when the last chunk is short). In all the datasets we studied, the last sentence or two consist of all or most of the query. Thus, the $p$ vector treats the query node(s) as the start nodes in the distribution of random walks modeled by the PPR distribution, biasing it toward items related to the query.

\textbf{Dynamic $\boldsymbol{\alpha}$.} The amount the PPR distribution is biased toward $p$ depends on the weight term $\alpha$. We hypothesize certain tasks require different levels of query-relatedness. Controlling $\alpha$ in our algorithm provides a straightforward method for controlling levels of query-relatedness during retrieval. We tested a wide range of $\alpha$ values one tasks that require local, query-dependent retrieval (e.g., QA and NIAH tasks) and tasks that require global, query-independent retrieval (e.g., summarization) (figure \ref{fig:alpha_test}). We found $\alpha$ values between $.5$ and $.99$ lead to high retriever recall accuracy across all query-dependent tasks, whereas $.1 \leq \alpha \leq .5$ performed slightly worse, and $\alpha = 0$ failed completely. However, on query-independent tasks, smaller $\alpha$ values performed better, and $\alpha = 0$ performed best. This justifies our hypothesis and suggests a simple $\alpha$-controller that acts as a router for a mixture of two experts: PPR with $.5 < \alpha$ for local, query-dependent tasks and PR, i.e., $\alpha = 0$ for global, query-independent tasks. We implemented a simple router that uses the base LLM with a zero-shot prompt. The LLM takes as input the last two and first two sentences of the context and judges whether ('y' or 'n') the task requires local or global retrieval. We found this method worked nearly perfectly across all the tasks we tested on (see figure \ref{fig:classify}). 

\begin{figure}[t]
\centering
\includegraphics[width=.92\textwidth]{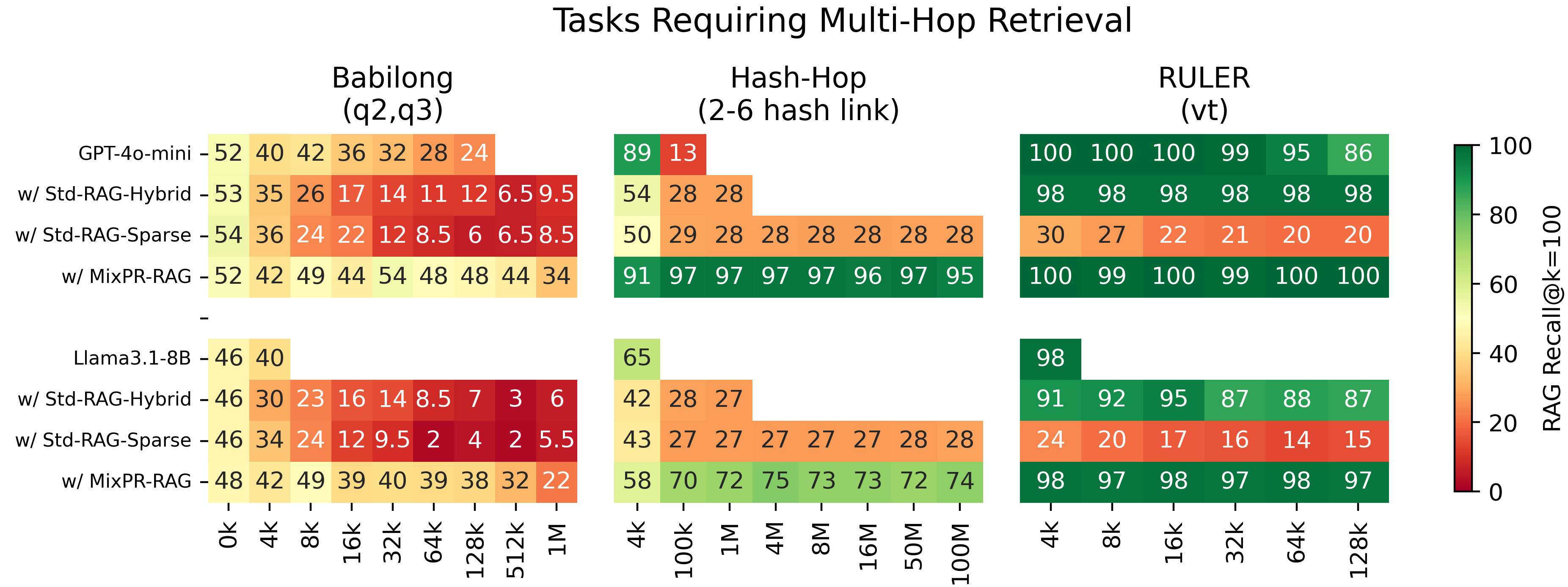}
\caption{\textbf{Performance on multi-hop retrieval tasks.} Results from benchmarks on the subset of tasks that involve multi-hop retrieval: BABILong question types 2 and 3, Hash-Hop with 2-6 hash links, and the variable tracing task from RULER. All RAG models tested with k=100. The non-graph baseline RAG models struggle, while the MixPR model is effective across all tasks improving performance over RAG baselines by as much as $50\%$. Note that although variable tracing is intended to require multi-hop retrieval \cite{hsieh2024ruler} the RAG-hybrid model is able to 'shortcut' this task in a single-hop retrieval step. See experiments section for discussion.}\label{fig:local_ret_multi}
\end{figure}
 
\textbf{A Fast, Efficient Implementation via Sparse Matrices.} In order to make RAG practical and advantageous over full context methods, our retrieval method must be memory efficient and fast. We achieve this through the use of sparse, keyword-based embeddings stored in sparse matrices implemented using largely standard Python libraries. We use the TF-IDF algorithm \cite{salton1988term}, implemented via the sklearn library \cite{pedregosa2011scikit}, to compute the sparse keyword-based embeddings. TF-IDF represents text chunks with word count vectors, which count the number of occurances of each word within a text chunk then weight this count (inversely) by the word's frequency across text chunks. As we show below, MixPR and its implementation embed text chunks much faster on a CPU than dense embedders do on a GPU. Embeddings are stored in sparse matrices, via the SciPy library \cite{2020SciPy-NMeth}. These matrices are highly memory efficient, only storing non-zero values, and the matrix multiplication used in our PPR retriever can be done fast, entirely on the CPU. We show below that chunking, embedding, and retrieving can be done in less than a few seconds, on all datasets up to 1M context. The fact that our retrieval method runs solely on CPU means that it can run in tandem with an LLM running on the GPU.

\section{Experiments}

We test our MixPR retriever across a variety of tasks and compare to well-established baselines. For comprehensive analysis, we separate sub-tasks into those that require local retrieval (i.e., high-query relatedness) from those that require global retrieval (high-structural importance). We further split local retrieval tasks into the sub-categories of single-hop and multi-hop tasks. Single-hop tasks are those in which relevant text chunks are directly related to query (e.g., keyword overlap, synonym sharing, etc.). Multi-hop retrieval tasks are those in which relevant text chunks may not be directly related to the query, but instead only have an indirect linking to the query through one or more other relevant text chunks (see appendix for examples). 

\textbf{Datasets.}
We run tests on four datasets. First, we test on the BABILong \cite{kuratov2024babilong} dataset. BABILong uses five questions from the BABIL \cite{weston2015babi} dataset, which require reasoning over a series of statements about people, places, things, and their interactions. It then uses the PG-19 dataset \cite{rae2019compressive}, a dataset of english language books, to add filler statements between the relevant sentences. Second, we use the RULER dataset \cite{hsieh2024ruler}. RULER consists of 13 tasks, including multiple needle-in-the-haystack (NiaH) tasks, 2 QA tasks, 2 'agglomerative' tasks which involve estimating word frequency over the whole input document, and 1 multi-hop reasoning task, called variable tracing. Third, we test on the Hash-Hop benchmark \cite{MagicHashHop}. Hash-Hop is a variable tracing task. A long list of hash assignments are presented, and the model must output all hashes equal to some query hash. Following earlier work \cite{MagicHashHop}, we test across 1-6 hash assignments. Finally, to better test tasks that require global retrieval, we also test on the English summary task from \cite{zhang2024infbench}, which involves summarizing entire english language books ranging from about $80$ to $~800$ thousand tokens. In total, we test on 22 tasks with 14 of these being single-hop local retrieval tasks, 8 multi-hop retrieval tasks, and 3 global retrieval tasks.
\begin{figure}[t]
\centering
\includegraphics[width=0.93\textwidth]{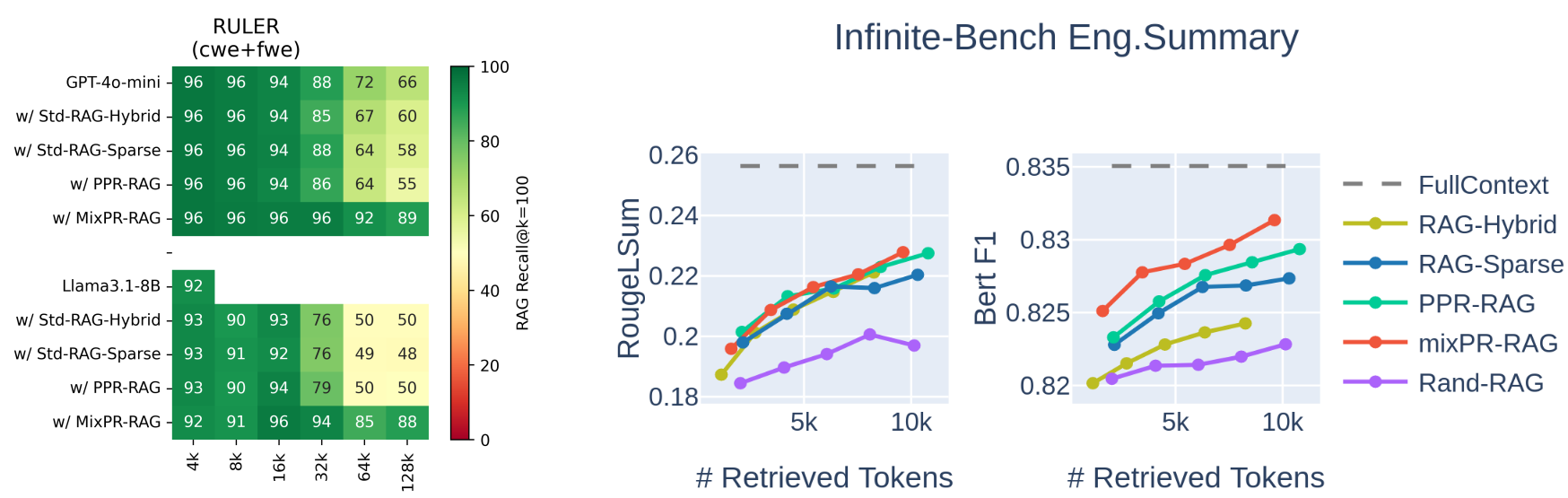}
\caption{\textbf{Performance on global retrieval tasks.} (Left) Performance of various RAG models with k=100 averaged across the cwe and fwe tasks from RULER. (Right) Performance of various RAG models on the Eng.Sum task from infinite-bench. Models test with k of 100, 200, 300, 400, 500.}\label{fig:global_ret}
\end{figure} 

\textbf{Models and Baselines.} We focus on comparing three different RAG methods across a range of base LLM scales, focusing on GPT-4o-mini \cite{hurst2024gpt} and Llama3.1-8B \cite{dubey2024llama}.  We compare our mixture-of-PageRanks retriever (\textbf{MixPR}) to two baseline retrievers. We compare our model with efficient RAG methods that may be practical for long-context tasks. SOTA graph-retrieval methods (e.g., \cite{peng2024graph,gutierrez2024hipporag}) use a very compute-expensive process to construct graph-databases, in which an LLM generates a list of related entities observed in the text, before embedding them into a database. This is much too inefficient and impractical for fast long-context processing. As such, we compare two simpler but faster RAG retrievers, based on those used in recent applications of RAG to long-context tasks (e.g., \cite{yu2024defense}). These retrievers use standard nearest neighbor-search, (i.e., retrieve top-k chunks with highest cosine similarity to query) and retrieve text, while maintaining the chronological order of text chunks. The first baseline, like our MixPR, uses TF-IDF \cite{salton1988term} sparse, keyword embeddings (\textbf{RAG-Sparse}) but performs only a standard (non-PPR-based) nearest neighbor search. The second baseline is a hybrid retriever (\textbf{RAG-Hybrid}) that uses a weighted average of similarity values from sparse, TF-IDF embeddings, and dense semantic embeddings from the SOTA 500M BERT-based embedder from Alibaba \cite{li2023towards}. Hybrid retrievers have been found to work better than either dense or sparse retrievers alone \cite{bruch2023analysis}. We give dense embedding similarity values a weight of .85, and sparse embeddings similarities a weight of .15. Our MixPR retriever uses the same $\alpha$ (of .6) and the same chunking procedure across all datasets.

\textbf{Alpha Testing.} We perform an isolated test of a PPR retriever (w/o LLM generation) across different $\alpha$ values. We split tasks into those that require local retrieval (retrieval of items with high query-relatedness) and global retrieval (retrieval of items with high structural importance). For local retrieval tasks we test recall at k=100. For global retrieval tasks, we test the retrieved text on the average of the cwe and fwe tasks and summarization. For cwe and fwe, we score the retriever by measuring the proportion of the top-5 most common words in the full context that are also in the top-5 most common words of the retrieved text. For summarization, we measure the RougeLSum of the retrieved text, which is the performance measure used in infinite-bench. We find (figure \ref{fig:alpha_test}) that recall for local retrieval tasks remains relatively constant across all $\alpha > 0$, with a small drop in performance on some RULER tasks when $\alpha < .5$, and a large drop in performance when $\alpha = 0$. In global retrieval tasks we observe a noticeable improvement in performance as $\alpha \rightarrow 0$. These results support our intuition that different tasks benefit from different $\alpha$ values, where more local retrieval tasks need larger $\alpha$, while global retrieval tasks need small $\alpha$ near 0.

\textbf{Local retrieval: single-hop tasks.} We compare MixPR to the baseline RAG models on local retrieval tasks. We test with k=100 for all RAG models. Results are in the appendix figure \ref{fig:local_ret_one}, top. As expected, we find all RAG models perform well. Importantly, we observe that across all tasks, RAG systems outperform the base long-context LLM at longer sequence lengths, showing that removing noise from the context on these tasks provides significant benefit to the LLM generator.

\begin{table}
  \centering
  \resizebox{\textwidth}{!}{\begin{tabular}{llllllllllll}
  \toprule
  \multicolumn{11}{c}{\textbf{BABILong Benchmark}}\\
  \toprule
    \textbf{Model} & $0k$ &  $4k$ & $8k$ & $16k$ &  $32k$  & $64k$ & $128k$ & $500k$ & $1M$ & $10M$ & Avg.\\
    \midrule
    ARMT (Finetune)  \cite{bab_leaderboard} & $99$ &  $98$  & $98$ & $98$ &  $98$  & $97$ & $96$ & $95$ & $93$ & $76$ & $\underline{94.8} \text{ (1st)}$ \\
    RMT-Retrieval (Finetune)  \cite{bab_leaderboard} & $98$ & $92$ &  $90$  & $87$ & $82$ & $73$ & $64$ & $48$ & $44$ & $32$ & $\underline{71.0}$ \text{(3rd)}\\
    RMT (Finetune) \cite{bab_leaderboard} & $99$ & $92$ & $89$ & $85$ & $77$ & $69$ & $58$ & $46$ & $42$ & $33$ & $\underline{69.0}$ \text{(4th)}\\
    \midrule
    GPT-4o & $92$ & $79$ & $75$ & $68$ & $63$ & $56$ & $50$ & - & -  & - & $48.3$\\
    \textbf{GPT-4o +MixPR-RAG}& $91$ & $85$ & $83$ & $86$ & $85$ & $83$ & $85$ & $81$ & $75$  & $50$ & $\underline{80.4}$ \text{(2nd)}\\
    \midrule
    GPT-4o-mini & $73$ & $66$ & $65$ & $59$ & $53$ & $52$ & $45$ & - & - & - & $41.3$\\
    \textbf{GPT-4o-mini +MixPR-RAG}& $73$ & $67$ & $71$ & $68$ & $72$ & $69$ & $69$ & $66$ & $61$ & $44$ & $\underline{66.0}$ \text{(5th)} \\
    \midrule
    Llama3.1-8B \cite{bab_leaderboard} & $67$ & $66$ & $62$ & $60$ & $56$ & $49$ & $39$ & - & - & - & $39.9$\\
    \textbf{Llama3.1-8B +MixPR-RAG}& $69$ & $62$ & $63$ & $60$ & $60$ & $58$ & $57$ & $54$ & $50$ & $45$ & $57.8$\\
    \midrule
    Llama3-ChatQA-1.5-8B +RAG \cite{bab_leaderboard}& $48$ & $46$ & $45$ & $45$ & $44$ & $42$ & $45$ & $42$ & $39$ & $37$ & $43.3$\\
    \bottomrule
  \end{tabular}}
\caption{\textbf{BABILong benchmark}. We compare MixPR RAG to the current top 4 models according to the BABILong leaderboard (Nov. 2024) \cite{bab_leaderboard}. The current top 3 models are variants of RMT architectures fine-tuned for BABILong specifically. The fourth is a RAG baseline from \cite{kuratov2024babilong} that uses simple nearest-neighbor search. Our RAG models perform much better than this baseline and are comparable to the fine-tuned models (being second only to ARMT \cite{rodkin2024associative}) especially at longer context sizes, despite not fine-tuning for this task and maintaining the full LLM capabilities. Top-5 underlined and marked.}\label{table:bab}
\end{table}

\textbf{Local retrieval: multi-hop tasks.} We find a clear difference, seen in figure \ref{fig:local_ret_multi}, between MixPR and the baseline RAG systems on multi-hop tasks. MixPR achieves high accuracy on all of these tasks consistently across input lengths, whereas the baseline RAG models struggle even at shorter lengths, and performance worsens quickly as the input length grows. One exception is the variable tracing task (vt) from RULER. The RAG-hybrid model is able to perform well on this task. A closer examination of the task shows that each variable assignment is labeled with 'VAR', which has a semantic relation to the word 'variable' in the query, allowing the semantic embedder to identify  variable assignment statements with a single similarity comparison. This suggests that this test is weak test of multi-hop retrieval (see also \cite{vodrahalli2024michelangelo} for similar critique). Finally, it is important to note that LLMs augmented with the MixPR-RAG perform better than the base LLMs alone, especially at longer text lengths. 

\textbf{Global retrieval tasks.} We test the models on three global retrieval tasks. The first two are from RULER: common words extraction (cwe) and frequency word extraction (fwe). Each task asks the model to report the top-N most common/frequent words. CWE increases the number of uncommon words with sequence length. FWE increases the number of occurrences of all words according to a Zeta distribution (word rank). We find consistently (Figure \ref{fig:alpha_test}) that MixPR (which uses $\alpha =0$ in these tasks) performs better than base LLM and baseline RAG models, including a PPR RAG model with $\alpha=.6$. The english summary task from infinite-bench \cite{zhang2024infbench} presents a book and a query asking the model to summarize the book. A human written reference summary is used for comparison. Infinite-bench uses rouge-l-sum as its measure. We also measure quality using BertF1 score, a neural network based metric that is more sensitive to semantics. We test across multiple k values. We find rouge scores are similar across RAG models except for the baseline of randomly sampling k chunks. However, we find Mix PR achieves a small but noticeable improvement over other RAG methods according to BertF1 score. It should be noted, however, that summarization quality is notoriously difficult to assess and scores like Rouge-L do not necessarily correlate well to human evaluations. Understanding the performance of MixPR-RAG for summarization and other global retrieval tasks could benefit from more accurate and nuanced assessment methodologies.

\begin{table}
  \centering
  \resizebox{.72\textwidth}{!}{\begin{tabular}{llllllll}
  \toprule
  \multicolumn{8}{c}{\textbf{RULER Benchmark}}\\
  \toprule
    \textbf{Model} & $4k$ &  $8k$  & $16k$ &  $32k$  & $64k$ & $128k$ & Avg.\\
    \midrule
    Jamba-1.5-large \cite{RULER_Leaderboard} & $96.7$ &  $96.6$  & $96.4$ & $96.0$ &  $95.4$  & $95.1$ & $\underline{96.0}$ \text{(2nd)}\\
    Gemini-1.5-pro \cite{RULER_Leaderboard} & $96.7$ &  $95.8$  & $96.0$ & $95.9$ &  $95.9$  & $94.4$ & $\underline{95.8}$ \text{(3rd)}\\
    Jamba-1.5-mini \cite{RULER_Leaderboard} & $95.6$ & $95.6$ &	$94.8$ & $94.6$ & $92.8$ & $90.0$ & $93.9$\\
    GPT-4-1106-preview \cite{RULER_Leaderboard} & $96.6$ & $96.3$ & $95.2$ & $93.2$ & $87.0$ & $81.2$ & $91.6$ \\
    \midrule
    GPT-4o* & $96.6$ & $96.6$ & $96.5$ & $96.1$ & $94.8$ & $91.1$ & $\underline{95.3}$ \text{(4th)}\\
    \textbf{GPT-4o +MixPR-RAG}& $96.9$ & $96.8$ & $96.6$ & $96.7$ & $96.1$ & $95.6$ & $\underline{96.5}$ \text{(1st)}\\
    \midrule
    GPT-4o-mini & $96.0$ & $95.3$ & $94.6$ & $93.3$ & $88.9$ & $85.4$ & $92.3$ \\
    \textbf{GPT-4o-mini +MixPR-RAG} & $96.1$ & $95.4$ & $95.1$ & $95.2$ & $94.1$ & $93.6$ & $\underline{94.9}$ \text{(5th)}\\
    \midrule
    Llama3.1-8B \cite{RULER_Leaderboard} & $95.5$ & $93.8$ & $91.6$ & $87.4$ & $84.7$ & $77.0$ & $88.3$ \\
    \textbf{Llama3.1-8B +MixPR-RAG} & $94.1$ & $93.5$ & $91.2$ & $91.9$ & $90.2$ & $89.3$ & $91.7$ \\
    \bottomrule
  \end{tabular}}
\caption{\textbf{RULER benchmark comparison}. We compare MixPR models to the top-4 models currently (November 2024) on the RULER leaderboard \cite{RULER_Leaderboard} and to the base LLMs used in our RAG models. We find that GPT4o, when augmented with our MixPR system achieves state-of-the-art performance on the RULER benchmark. $^*$Note we test GPT-4o on a subset of RULER due to high costs.}\label{table:ruler}
\end{table}

\begin{table}
  \centering
  \resizebox{.9\textwidth}{!}{\begin{tabular}{lllllllllll}
  \toprule
  \multicolumn{10}{c}{\textbf{Hash-Hop Benchmark}}\\
  \toprule
    \textbf{Model} & $100k$ &  $200k$  & $1M$ &  $2M$  & $4M$ & $8M$ & $16M$ & $50M$ & $100M$ & $500M$\\
    \midrule
    LTM-Magic.dev (Finetune) \cite{MagicHashHop} & $100$ &  $100$  & $100$ & $100$ &  $100$  & $100$ & $100$ & $95$ & $90$ & -\\
    \midrule
    \textbf{GPT-4o-mini +MixPR-RAG} & $97$ & $97$ & $97$ & $97$ & $97$ & $97$ & $96$ & $97$ & $96$ & $97$\\
    \textbf{Llama3.1-8B +MixPR-RAG} & $75$ & $78$ & $76$ & $78$ & $79$ & $77$ & $77$ & $77$ & $78$ & $77$\\
    \bottomrule
  \end{tabular}}
\caption{\textbf{Hash-hop benchmark comparison} averaged across tests of 1-6 hop lengths.}\label{table:HH}
\end{table}

\textbf{Comparisons to SOTA.} In tables \ref{table:bab} \ref{table:ruler} \ref{table:HH}, we compare MixPR to the current state-of-the-art on the RULER, BABILong, and Hash-Hop benchmarks. There are two main observations. First, consistent with the our other results above, adding MixPR improves performance over the base LLM. The average score achieved by GPT-4o-mini on RULER improves by 2.6 points when augmented with MixPR. Llama3.1-8B improves by 3.4 points. More drastic improvements can be observed on Babilong results. The second observation is that our RAG models are able to compete with or improve upon SOTA results. For example, GPT-4o w/ MixPR is second on the BABILong leaderboard and first on RULER. GPT-4o-mini w/ MixPR is fourth on BABILong and third on RULER. Among RAG models thus far tested on BABILong, our MixPR model performs the best, improving the previous baseline by 20+ points. On Hash-Hop, GPT-4o-mini w/ MixPR all but matches the SOTA model from Magic.dev \cite{MagicHashHop}, which is fine-tuned on Hash-Hop, and outcompetes it by 6 points at 100M token length.

\begin{figure}[t]
\centering
\includegraphics[width=.825\textwidth]{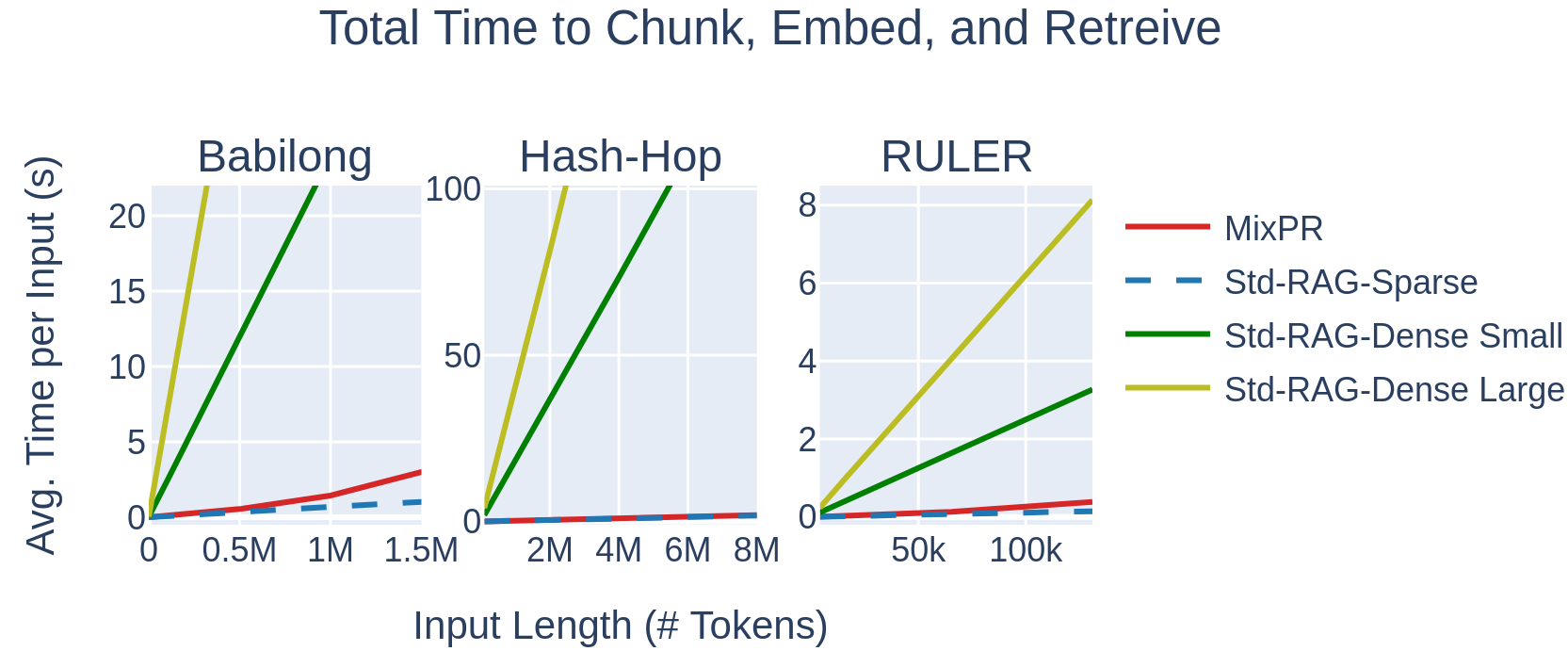}
\caption{\textbf{RAG compute times in compute constrained setting.} Total time for RAG systems to chunk, embed, and retrieve from a long context. Retrieval systems are run on hardware available in high-end desktops: CPU with 16 cores and a single RTX 4090 GPU. We compare our MixPR retriever with the sparse retriever that does nearest neighbor search via FAISS flat search (RAG-Sparse) and two dense retrievers. We test a dense retriever that uses a 500M parameter BERT-style embedder (RAG-Dense Large) and a 30M parameter BERT-style embedder (RAG-Dense Small). }\label{fig:compute_infer}
\end{figure}

\textbf{Compute cost analysis.} In figure \ref{fig:compute_infer}, we show the amount of time it takes for various retrievers to chunk, embed, and retrieve from the context at different sequence lengths. We measure retrieval time in a setup with a CPU and GPU available in a high-end desktop, an AMD Ryzen Threadripper PRO 5955WX CPU with 16-Cores and a single RTX 4090 GPU. We compare MixPR to RAG-Sparse and as well as to two dense retrievers. The MixPR and RAG-Sparse models use a symbolic program to create TF-IDF embeddings. The first dense retriever uses a 500M parameter BERT embedder, while the second uses a 33M BERT embedder. These dense embedders are run on a RTX 4090 with batch size optimized for wall-clock time. All retrievers use the same chunking method. We observe the sparse models are very fast, taking only a few seconds to process even 1.5M+ tokens, while the dense embedders are much slower, taking several minutes to process the same amount of tokens. The main bottle-neck is embedding time: the dense embedders can be slow to iterate though all mini-batches of text chunks on the smaller GPU we use, while the program that computes TF-IDF embeddings runs with extremely low latency on the CPU.

\section{Conclusion}
In this paper, we presented our mixture-of-pageranks (MixPR) retriever. We demonstrate that this retrieval algorithm outperforms existing RAG methods and also long-context frontier LLMs on a wide range of challenging long-context benchmarks. Our method, when used to augment GPT4o models, achieves SOTA results on BABILong, HashHop, and RULER. It also sometimes can match or outperform the performance of specialized retrieval architectures fine-tuned for these benchmarks.

Moreover, our MixPR method is exceptionally compute efficient due to using sparse embeddings for retrieval. This enables our method to run effectively in real-time on the CPU, and hence can run in tandem with an LLM system on the GPU. This is in strong contrast with both existing RAG systems which use neural network retrievers which are both expensive and must run on the GPU, as well as directly inferencing a long-context model at its full context-length which can also be extremely expensive and somewhat slow. Our MixPR system could thus be a promising basis for making long-context systems much more economical and ubiquitous.

\bibliographystyle{plain}
\bibliography{references}


\appendix

\section{Appendix}

\subsection{Extended Related Works}

\textbf{Long-Context LLMs.} LLMs with long-contexts (e.g., \cite{team2024gemini, anthropic2023model, achiam2023gpt}) have recently seen significant development and attention. While these models have shown promising results on various benchmarks (e.g., \cite{RULER_Leaderboard}), they typically utilize transformer attention layers, which means these models can have enormous computational costs, making them impractical for compute-constrained scenarios.

\textbf{Long-Context Benchmarks.} Benchmarks testing the ability of these LLMs to effectively utilize their context are currently being rapidly developed. Early long context benchmarks focus on retrieval tasks, especially synthetic needle-in-the-haystack (NIAH) tasks \cite{kamradtNIAH, team2024gemini, hsieh2024ruler, li2024needlebench, zhang2024infbench}. Similar retrieval tasks have also been extended to natural language settings, e.g., \cite{kuratov2024babilong, hsieh2024ruler, zhang2024infbench}. More recently, there has been an emphasis on testing more complex tasks than mere single step retrieval \cite{levy2024sametask, goldman2024really, karpinska2024one, vodrahalli2024michelangelo}. The RULER benchmark \cite{hsieh2024ruler} and the HashHop benchmark \cite{MagicHashHop}, for example, develop variable tracing tasks, which requires transitive reasoning over variable assignments dispersed in a larger text. Other benchmarks incorporate reading comprehension \cite{karpinska2024one}, summarization \cite{zhang2024infbench}, and word frequency estimating tasks \cite{hsieh2024ruler}, which favor analyzing the input globally. Vodrahalli et al. \cite{vodrahalli2024michelangelo} recently presented three difficult long-context reasoning tasks which they claim require LLMs to utilize the "latent structure" within the input document, while ignoring the structurally irrelevant portions. This dataset proved difficult for SOTA long context LLMs, but the dataset has yet to be released publicly.

\textbf{Graph-based RAG for Database Retrieval} Graph-based RAG systems retrieve from graph structured databases rather than unstructured databases. Edge et al. \cite{edge2024local} show that graph structured databases may be used to better respond to queries about the global properties of the database, rather than local, query-specific properties. Gutierrez et al. \cite{gutierrez2024hipporag} get SOTA results on multi-hop QA tasks using a graph-structured database and the personalized page rank retrieval algorithm. Other recent results suggest graph-based RAG is promising relative to unstructred database retrieval (for review, see \cite{peng2024graph}). We take inspiration from these prior works. However, we utilize different methods for creating the graph-structured database that allow for fast text embedding and graph construction that can occur at test time, whereas previous works use pre-embedded knowledge graphs or use an expensive graph construction process that utilizes LLMs to process and output entity-relation triplets, and semantic neural network embedders.

\textbf{RAG for Long Contexts.} Alonso and Millidge \cite{AlonsoMillBabilongBlog} tested several graph-based RAG systems on the Babilong \cite{kuratov2024babilong}, long-context QA task. They showed chronological ordering of retrieved chunks and graph-based retrieval is necessary to achieve good performance on difficult multi-hop questions. Yu et al. \cite{yu2024defense} subsequently showed that standard RAG systems perform well on two simple long-context QA tasks and provided further evidence chronological ordering of text chunks is important. Zhao et al. \cite{zhao2024longrag} developed a complex RAG system that performs multiple retrieval steps guided by a chain of thought process, and showed it was effective on QA tasks. Unlike the work in this paper, none of these previous works attempt to address the main challenges for RAG-based long context models, as they do not test on complex reasoning or summarization tasks, nor do they analyze compute costs of RAG or attempt to minimize the compute cost in any direct way.

\textbf{Alternative Architectures} Language models based on state-space models such as Mamba \cite{gu2023mamba} and RWKV, as well as specialized retrieval architectures such as recurrent memory transformers \cite{bulatov2022recurrent, kuratov2024babilong, rodkin2024associative}, require substantially less memory and compute than transformers and have shown some favorable results when fine-tuned for a specific long-context tasks \cite{kuratov2024babilong, rodkin2024associative}. Similar architectures, like the auto-compressor \cite{chevalier2023adapting} and transformer with Beacon tokens \cite{zhang2024longbeacon} transformers, have shown some promising results on long-context language modeling tasks. However, there is evidence such recurrent models cannot generalize to long-context tasks they have not been fine-tuned for, e.g., Mamba-130M fine-tuned for BABILong performs very well, but pre-trained Mamba-2.8B completely fails \cite{kuratov2024babilong}. Moreover, such specialized model architectures often perform poorly at general language modeling. For example, the auto-associative RMT (ARMT) model \cite{rodkin2024associative} has the highest score on BABILong, but performs significantly worse than Mamba and GPT2 on Wiki-text language modeling (see appendix \cite{rodkin2024associative}). Other work has explored sparse attention mechanisms to reduce the compute cost of the dense, softmax attention of standard transformers with some success on moderately long context tasks (e.g., \cite{xiao2024infllm, fountas2024episodic, klett2024extended}). However, such methods sometimes do not match RAG performance \cite{klett2024extended}, have not been rigorously tested on long-context tasks with 100k+ tokens, and can require significant architectural modifications.

\subsection{Methods}

\textbf{Implementation Details for MixPR.} After computing the embeddings, E, using TF-IDF, these are L2 normalized then multiplied to get: $A = E^{\top}E$. We further sparsify the adjacency matrix A such that it has over 50 million non-zero  entries, by removing all values below .27. We then use a similar implementation of PPR to that implemented by networkx libary \cite{hagberg2008networkx}, except our implementation has a max number of iterations (which we set to 18) it will perform before outputting the distribution, whereas networkx throws an error when max iterations are reached. We set $\alpha=.6$ for the PPR expert in the MixPR. 

To classify the query we extract the first and last two text chunks of the input, which we use as the\texttt{[query]} variable in the prompt below. We prompt the RAG model's LLM generator with the following, which directs it to classify the query:

\colorbox{prompt_gray}{\begin{minipage}{.95\textwidth}
\texttt{If the query asks for summarization, most frequent words, or descriptions of an entire document, answer 'y' with no other text. 
If the query does not and only asks a specific question, answer 'n' with no other text.}
\newline

\texttt{QUERY: [query]}  
\newline

\texttt{If the query asks for summarization, most frequent words, or descriptions of an entire document, answer 'y' with no other text. If the query does not and only asks a specific question, answer 'n' with no other text.}
\newline

\texttt{Answer:}
\end{minipage}}

\textbf{Implemention Details for Baseline Models} RAG-Sparse uses the same procedure as the MixPR to get normalized TF-IDF embeddings of the context, stored in sparse scipy matrices. It then does a simple cosine similarity search to get the similarity values between the query vector, $q$, a TF-IDF representation of the last sentence (or two if the last sentence is fewer than 3 words). This is achieved through a simple vector matrix multiply performed multiply on CPU: $s_{sparse} = E^{\top}q$. 

RAG-Hybrid uses the same method as RAG-Sparse to get sparse, keyword based similarity scores, but it also combines these with dense embedding-based similarity scores. It uses the 435M Alibaba bert-style embedder \cite{li2023towards}, which at the time of writing was SOTA in its parameter class, to get dense embeddings. We processed all text chunks with mini-batch size 16.Embeddings are L2 normalized and cosine similarity scores are computed by using an FAISS \cite{douze2024faiss} flat inner product index and its built in retrieval operation. The final similarity scores are $s_{hybrid} = .15  s_{sparse} + .85  s_{dense}$, which we found through a grid search.

\textbf{BABILong Tests} We download the BABILong dataset from HuggingFace. To evaluate, we use the evaluation function in the BABILong code base \cite{kuratov2024babilong}. We use essentially the same prompts as used in the baseline RAG models test on BABILong, except, importantly, we concatenated the prompts to the input text, so the RAG system must retrieve from the prompt as well as the document. This makes the task harder for the RAG model compared to retrieving from the document only then concatenating the full prompt afterwards. However, we believe this is more realistic (i.e., one typically inputs a document concatenated with the prompt and query all at once into LLMs). 

\textbf{RULER Tests} We use the RULER codebase to generate datasets for each sub-task at each length. We generate and test on 500 questions for each sub-task, as in the original paper \cite{hsieh2024ruler}, with the exception of GPT-4o, which we only test on a subset due to very high costs of running. In particular, we find there is very little variance in the averaged scores after about 75 question on all tasks except QA, which varies more significantly as testing progresses. Consequently, we test GPT-4o on 75 questions from all tasks, except QA tasks where we test on 300 questions each. We use the same evaluation function as in the publicly available codebase to test our model outputs. Unlike BABILong, prompts are already concatenated to the input document and query in the dataset.

\textbf{Hash-Hop Tests} We use the Hash-Hop codebase to generate datasets for each hop-number and each token length. We generate 100 questions for each length. In order to get this to work for LLMs not fine-tuned on the task, we generate a simple few-shot prompt to instruct the LLM how to answerthe question. To evaluate, we check the proportion of hashes equal/assigned to the query hash that are present in the LLM output.

\subsection{Supplementary Results}

\begin{figure}[h]
\centering
\includegraphics[width=.92\textwidth]{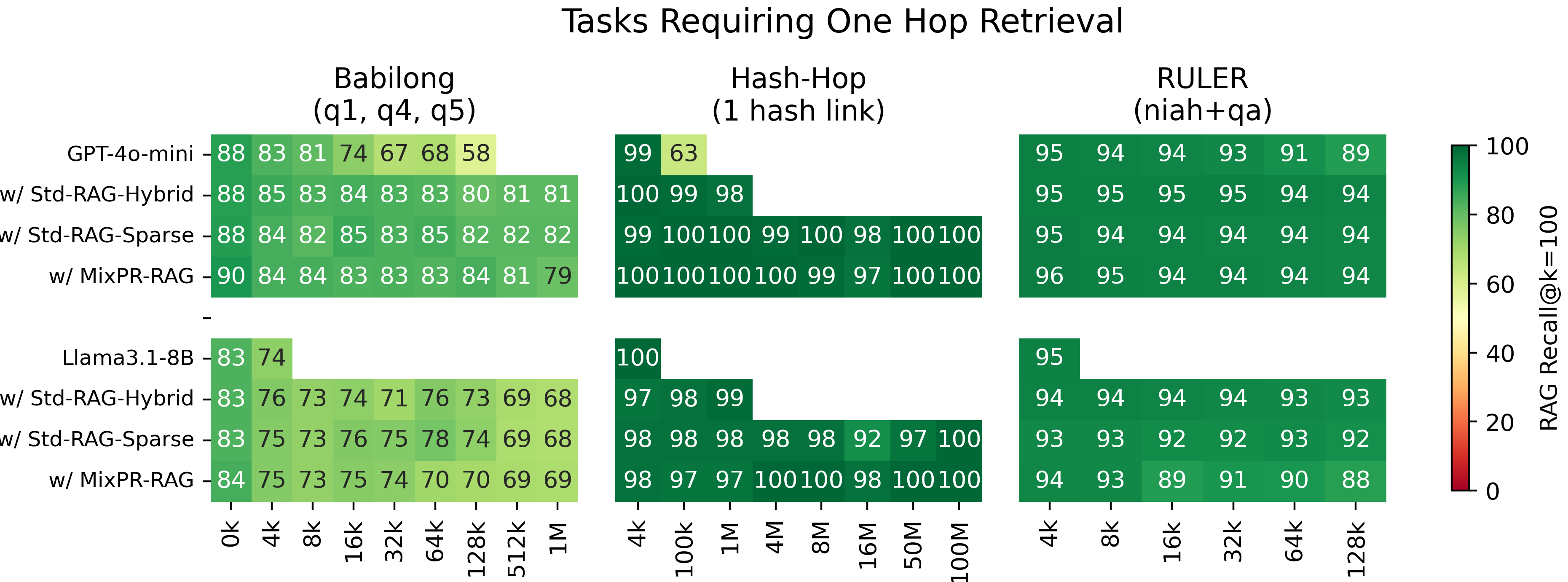}
\caption{\textbf{Performance on one-hop retrieval tasks.} Results on the subset of tasks that require one-hop retrieval: BABILong question types 1, 3 and 5, Hash-Hop with 1 hash link, and the needle-in-the-haystack and QA tasks from RULER. All RAG models tested with k=100. All models perform well, with RAG models matching or outperforming base models.}\label{fig:local_ret_one}
\end{figure}

\begin{figure}[h]
\centering
\includegraphics[width=.99\textwidth]{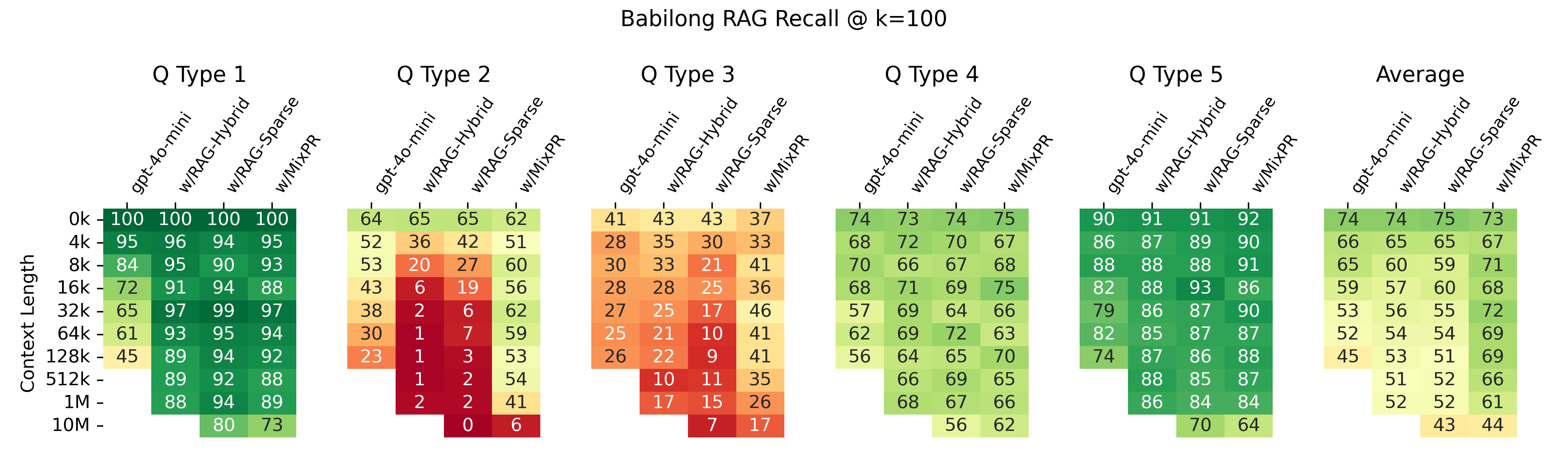}
\caption{BABILong Full Results GPT-4o-mini.}\label{fig:bab_gptmini_full}
\end{figure}

\begin{figure}[h]
\centering
\includegraphics[width=.99\textwidth]{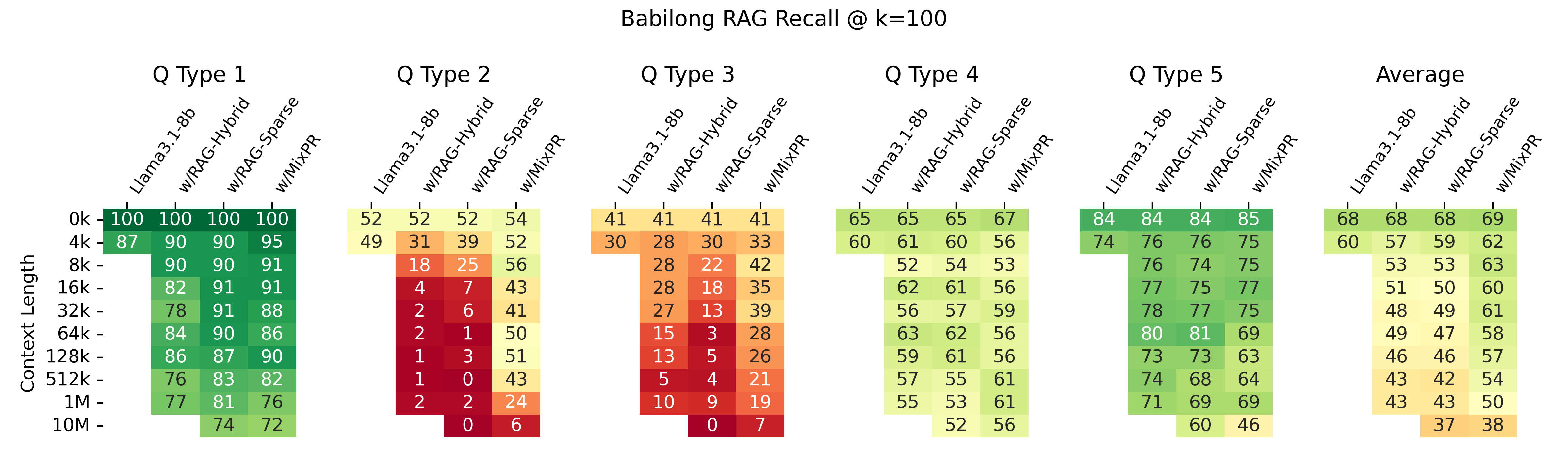}
\caption{BABILong Full Results Llama3.1-8B.}\label{fig:bab_llama_full}
\end{figure}

\begin{figure}[h]
\centering
\includegraphics[width=.8\textwidth]{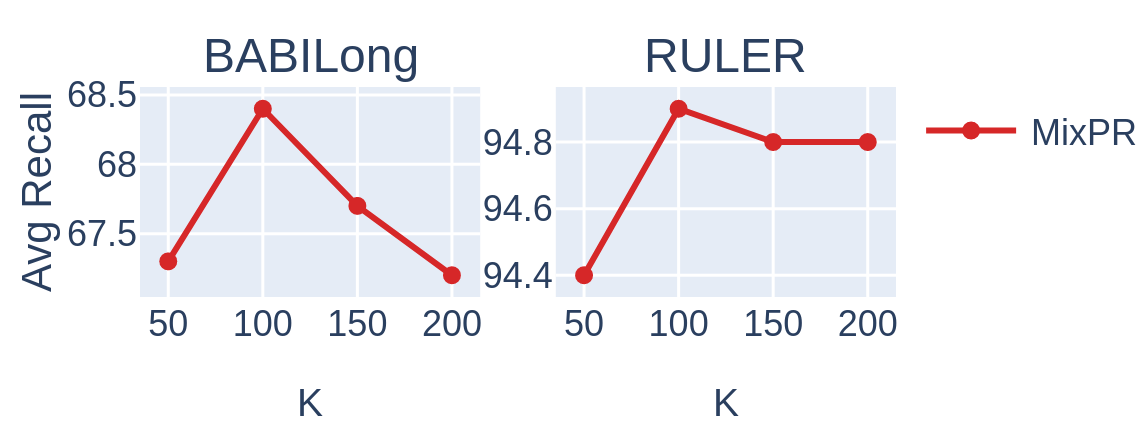}
\caption{\textbf{K-Analysis.} Test different K values for MixPR on RULER (all lengths) and BABILong (0k to 1M) using gpt-4o-mini as the LLM generator. For these tests, which consist mainly of local retrieval tasks, there is a sweet spot around k=100, after which noise/irrelevant text hurts performance and before which the retriever struggles to retrieve relevant text chunks.}\label{fig:k_analyze}
\end{figure}

\begin{figure}[h]
\centering
\includegraphics[width=.4\textwidth]{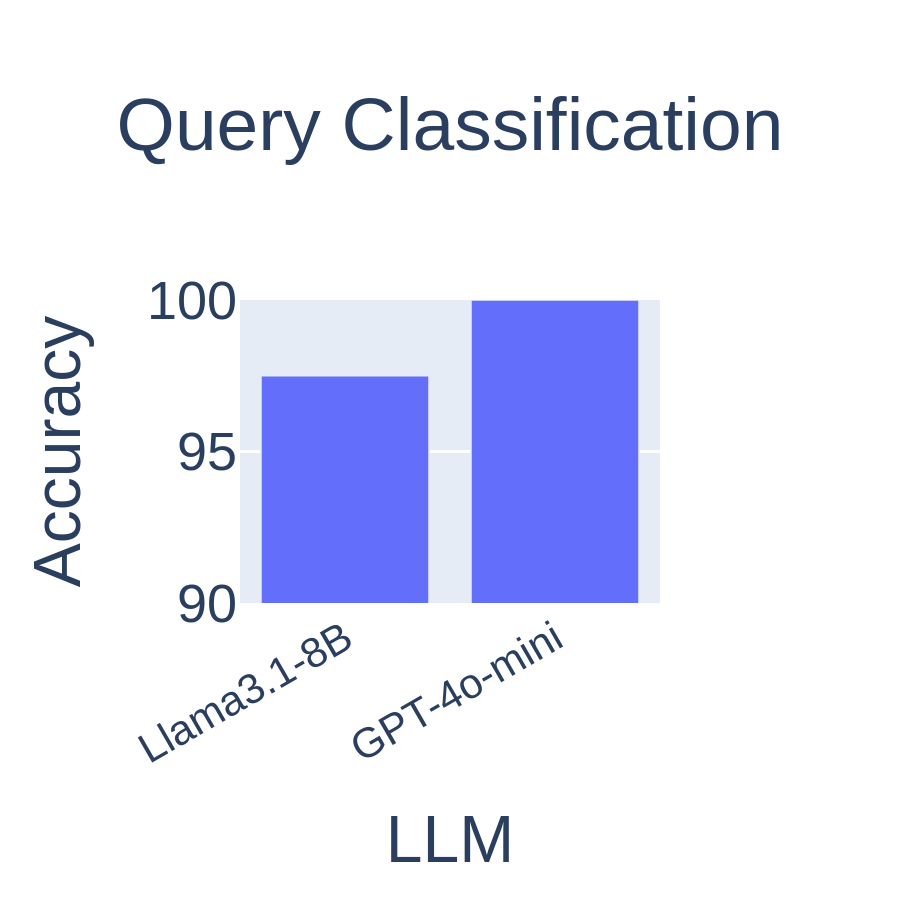}
\caption{\textbf{Classification Accuracies.} The MixPR RAG model uses the base LLM to classify queries as either requiring global or local retrieval using a prompting method. Here we show the classification accuracy across queries from every sub-task (22 total) from the benchmarks in the paper. For each task, there are 1 to 2 queries we test the classifier on totaling 38 questions. We show accuracy for Llama3.1-8B and GPT-4o-mini. GPT-4o-mini classifies perfectly. Llama3.1 only gets one question wrong, a summarization question.}\label{fig:classify}
\end{figure}
\end{document}